\documentclass[12pt]{article}
\usepackage{latexsym}
\usepackage{epsfig,graphics}

\newcommand{\be}{\begin{equation}}
\newcommand{\ee}{\end{equation}}

\def\lsi{\raise0.3ex\hbox{$<$\kern-0.75em\raise-1.1ex\hbox{$\sim$}}}
\def\gsi{\raise0.3ex\hbox{$>$\kern-0.75em\raise-1.1ex\hbox{$\sim$}}}

\begin{document}

\begin{titlepage}

\null\vspace{-1.0cm}

\begin{tabbing}
\` Edinburgh  2000-21 \\
\` Oxford OUTP-00-42P \\
\` Tsukuba UTCCP-P-93 \\
\end{tabbing}

\begin{centering} 

\null\vspace{1.0cm}

{\bf 
${\bf Z_2}$ MONOPOLES IN D=2+1 SU(2) LATTICE GAUGE THEORY}

\vspace{1.0cm}

A. Hart$^{{\rm a}\ast}$, B. Lucini$^{\rm b}$, 
Z. Schram$^{\rm c,d}$ and M. Teper$^{\rm b}$

\vspace{0.6cm}
{\em $^{\rm a}$%
Department of Physics and Astronomy, University of Edinburgh,\\
Edinburgh EH9 3JZ, Scotland, UK\\}
\vspace{0.3cm}
{\em $^{\rm b}$%
Theoretical Physics, University of Oxford, 1 Keble Road, \\
Oxford OX1 3NP, England, UK\\}
\vspace{0.3cm}
{\em $^{\rm c}$%
Department of Theoretical Physics, University of Debrecen, \\
H-4010 Debrecen P.O.Box 5, Hungary\\}
\vspace{0.3cm}
{\em $^{\rm d}$%
Center for Computational Physics,University of Tsukuba, \\
Tsukuba, Ibaraki 305-8577, Japan\\}

\vspace{1.25cm}
{\bf Abstract.}
\end{centering}

\noindent
We calculate the Euclidean action of a pair of $Z_2$ 
monopoles (instantons), as a function of their spatial 
separation, in D=2+1 SU(2) lattice gauge theory. 
We do so both above and below the deconfining transition 
at $T=T_c$. At high $T$, and at large separation, we find 
that the monopole `interaction' grows linearly with distance:
the flux between the monopoles forms a flux tube (exactly like
a finite portion of a $Z_2$ domain wall) so that the monopoles are
linearly confined. At short distances the interaction is
well described by a Coulomb interaction with, at most, 
a very small screening mass, possibly equal to
the Debye electric screening mass. At low
$T$ the interaction can be described by a simple screened Coulomb
(i.e. Yukawa) interaction with a screening mass that can be 
interpreted as the mass of a `constituent gluon'. None of this is 
unexpected, but it helps to resolve some apparent controversies 
in the recent literature.

\vfill

\noindent{$^\ast$ Current address: DAMTP, University of
Cambridge, Cambridge CB3 0WA, UK}
\end{titlepage}

\section{Introduction}
\label{sec_intro}

In a recent paper
\cite{rebbi}
the free energy of a separated pair of $Z_2$ monopoles has been
calculated in D=3+1 SU(2) 
lattice gauge theory. This problem is difficult because it
involves the numerical evaluation of the ratio of two different
partition functions, so that the calculations in
\cite{rebbi}
are only
accurate at small monopole separation. The conclusion is that
the interaction  potential has a simple Coulomb/Yukawa form 
both at high and at low $T$; and that this remains so,
at high $T$, even if one considers monopoles whose world-lines 
are spacelike rather than timelike. While the other results are
consistent with expectations, this last is not. What one expects
is that the $Z_2$ flux between the monopoles
will form a `flux tube' so that the
interaction energy grows linearly at large separation.
Motivated by this unexpected result we have performed
similar calculations using a different method that
is less direct but much more accurate. This method has been used
recently 
\cite{vor3,vor4}
in closely related investigations of vortices
\cite{thooft}
and, some time ago, of high-$T$  $Z_2$ domain walls
\cite{wall3}.
Our calculations are carried out in 2+1 Euclidean dimensions
where, apart from the fact that the monopoles are instanton-like
rather than soliton-like, not only can the same questions be 
posed as in 3+1 Euclidean dimensions but they can be answered
much more accurately. In addition the close relationship between
$Z_2$ monopoles, vortices and domain walls is particularly 
transparent in 2+1 dimensions. (Although it holds in D=3+1
equally well.) As we shall emphasise, after a suitable relabelling 
of the space-time axes, one can see 
\cite{vor3,wall3}
that the high-$T$ $Z_2$ domain wall is a periodic vortex line
(as created by a periodic 't Hooft disorder loop
\cite{thooft})
since this phase is a Higgs phase for vortices. And in this
phase the $Z_2$ flux between a pair of well-separated
monopoles passes through a flux tube that is nothing
but a portion of such a domain wall that ends on these
monopoles; so that such $Z_2$ monopoles 
are `linearly confined'. We shall indeed confirm
that the interaction is linear at high $T$, as expected. 
The contrary conclusion in
\cite{rebbi}
can presumably be attributed to the lack of accurate calculations
at larger inter-monopole separation. We have not pursued the
same calculations in D=3+1 since, as our D=2+1 calculations
were completed,  a paper appeared 
\cite{deforcrand}
addressing similar questions in D=3+1 with conclusions similar
to ours (where they overlap).

In the next section we describe $Z_2$ monopoles and explain
how we expect them to behave. We then describe our Monte Carlo 
calculations and finish with some conclusions. Since most of the
techniques and physical quantities involved have been 
described in detail in earlier papers
\cite{rebbi,vor3,wall3},
we shall aim to be succinct rather than pedagogical in this
brief report.

\section{${\bf Z_2}$ monopoles - expectations}
\label{sec_mono}

In 2+1 Euclidean dimensions, the monopoles are localised
in space and time and so can equally well be thought
of as instantons. What we shall do is calculate
the extra Euclidean action, $\Delta S$, associated with the 
presence of a pair of such monopoles. If we were in D=3+1
this quantity would be proportional to the derivative
of the free energy of the monopole pair, and its
variation with the monopole distance would tell us how the
monopole interaction energy varies with distance.
In D=2+1 we shall, by analogy, sometimes refer to the 
action difference as the monopole `interaction energy'.

To calculate the interaction between $Z_2$ monopoles at sites
$x$ and $x^\prime$ (on the dual lattice) one introduces a 
$Z_2$ Dirac string that starts at $x$ and ends at $x^\prime$
\cite{susskind}.
Such a string flips the sign of any plaquette that it
threads. (This is so if, as here and in the related literature, 
the SU(2) group elements on the links are represented by fundamental 
$2 \times 2$ matrices rather than by adjoint $3 \times 3$ matrices.)
One can introduce such a Dirac string by changing $\beta \to -\beta$
on the plaquettes threaded by the string. We shall refer to this
as a set of `twisted' plaquettes because, in the case where one
obtains $x^\prime$ by translating $x$ by a single lattice period,
this is equivalent to the imposition of twisted boundary conditions. 
By redefining suitable subsets of link matrices by 
$U^\prime = -U$ 
\cite{rebbi,vor3,wall3},
it is easy to see that the actual location of the string, apart from
its end points, is invisible; as we would wish for a Dirac string. 

\subsection{${\bf T > T_c}$}
\label{subsec_highT}

Let us first consider the extreme case
where the line of twisted plaquettes extends right across
the lattice; say in the $x$-direction with the twisted 
plaquettes in 
the $(y,t)$ plane. This produces a $Z_2$ Dirac string that closes
upon itself through the boundary; so it is not attached to
any monopoles. Such a twist is, as remarked earlier, equivalent
to imposing twisted boundary conditions. For $T > T_c$ it
imposes the presence of a $Z_2$ domain wall separating the
two high-$T$ $Z_2$ vacua
\cite{wall3}.
As pointed out in 
\cite{vor3,wall3},
this domain wall is in fact a periodic spacelike vortex tube that
has been squashed by the narrow extent in the Euclidean
time direction. Its existence demonstrates that such spacelike
$Z_2$ flux is, at high $T$, confined to a flux tube.
So we expect that for  $T > T_c$ the interaction energy
between two spacelike separated $Z_2$ monopoles will grow linearly
with their separation, $r$, once this separation is larger than
the width of the tube
\be
\Delta S \simeq {\bar \sigma} r \ \ \ \ \  \ \ \ \ r \ \ {\rm large}.
\label{eqn_DS_rlarge}
\ee
(Note that all quantities will be expressed in lattice units
from now on.) At small separations the flux of one
monopole spreads out in the three Euclidean space-time directions
and then returns to the other monopole with no opportunity to form a
flux tube. Thus the interaction should be Coulombic $\propto 1/r$.
As $r$ is increased we might expect a Yukawa-type correction
\be
\Delta S \propto {{e^{-\mu r}}\over{r}} \ \ \ \ \ \ \ \ \ r \ \ {\rm small}.
\label{eqn_DS_rsmall}
\ee
with the scale $\mu$ being plausibly related to the Debye 
electric screening mass 
\cite{wall3}.
The latter governs the width of 
the vortex tube/domain wall which, for large $r$, carries the
return flux between the monopoles over most of the distance
between them. 

Note that if there were no linear term then the asymmetric
geometry of our system would mean that, as we increased 
$r$ to $r > 1/T$, any $1/r$ Coulomb term would turn into a 
logarithm $\propto \log(rT)$. This is also the case
if we consider spatially separated $Z_2$ monopoles at
high $T$ in 3+1 dimensions. We shall ignore this for
the purposes of our fits because, in practice, the Coulomb
or Yukawa contribution becomes very small for $r > 1/T$.

At very high T we expect the tension of the $Z_2$ domain wall to
be calculable in perturbation theory and, as shown in
\cite{wall3},
the perturbative expression is accurate down to remarkably small
values of $T/g^2$. (Recall that in D=2+1 $g^2$ has dimensions of
mass.) Thus the fact that our spacelike separated monopoles
are joined by a section of such a domain wall, allows us
to estimate the value of ${\bar \sigma}$ in eqn(\ref{eqn_DS_rlarge}):
\be
{\bar \sigma} = 2\alpha \biggl({ T\over g^2}\biggr)^\frac{3}{2} 
{1 \over \beta^2}
\label{eqn_sigma}
\ee
where $\beta$ is the usual (inverse) lattice coupling and
\be
\alpha = \alpha_{pert} (1+0.25{{g^2}\over{T}})
\label{eqn_alpha}
\ee
provides an accurate representation of the Monte Carlo
calculated domain wall tension for $g^2/T < 1.2$. The
leading order perturbative value, $\alpha_{pert}$, can
be calculated as a function of the temporal lattice
discretisation and the resulting values are listed in Table 8 of
\cite{wall3}.

In addition to the above quantitative expectation for
the interaction between spatially separated monopoles at
high $T$, one might equally well consider the interaction
of two monopoles with a timelike separation. However, because
such a separation is limited to very small values,
$r \leq 1/2T$, there is nothing very interesting to be learned 
and we shall not consider that case any further here.

\subsection{${\bf T < T_c}$}
\label{subsec_lowT}

Consider now the low-$T$ confining phase. In this case if
there are vortex tubes, they condense into the vacuum
and thus the interaction between monopoles becomes 
independent of $r$ at large $r$. At small $r$ we again
expect a screened Coulomb interaction $\propto \exp(-\mu r)/r$,
or a sum of such terms. But now there is no $T$ to set 
the scale, so (the lightest) $\mu$ 
will presumably be related to the mass gap, which is
provided by the lightest $0^{++}$ glueball in this theory
\cite{spectrum3}.

\vskip 0.3in

The above discussion translates directly to D=3+1. Now the
$Z_2$ monopoles will have world lines and the Dirac string will
sweep out a sheet, which requires a sheet of twisted plaquettes.
If we are at high $T$ and the monopole world lines
and separations are spacelike, then the return flux once
again goes through flux tubes,  with the scale set by
the electric Debye mass. So at large separations the interaction
energy should grow linearly and at small distances it
will be screened Coulombic. By contrast, at low $T$ the
vortex tubes condense in the vacuum and there is only
a screened Coulombic interaction. While the calculations of
\cite{rebbi}
agree with the above expectations at low $T$ they claim
that a simple Coulomb interaction, with no
linear piece, works at high $T$ as well. Our method of using
the action difference is, however, much easier and more 
accurate, especially at larger $r$, than the method 
employed in
\cite{rebbi}
and we now turn to calculations designed to resolve this
puzzle.

\section{${\bf Z_2}$ monopoles - calculations}
\label{sec_calc}

Our Monte Carlo calculations use the usual plaquette
action parameterised by the (inverse) coupling $\beta$
where $\beta \to 4/ag^2$ as $a \to 0$. The low-$T$
physics of this theory is discussed in 
\cite{spectrum3},
the properties at high $T$ in
\cite{wall3},
and the deconfining transition in 
\cite{Tc3}.

We perform calculations at two values of the coupling:
$\beta = 9$ and $\beta = 13.5$. In each case we perform
one calculation at $T > T_c$ and one at $T < T_c$.
At $\beta=9$ we perform calculations on $24^2 4$ and
$24^2 8$ lattices
and at $\beta = 13.5$ on $36^2 6$ and  $36^2 12$ lattices.
On a $L^2 L_t$ lattice of spatial extent $L \gg L_t$
we have $aT=1/L_t$ so the above choices correspond
to higher and lower $T$ respectively. To be more
quantitative we recall that the critical value of
$L_t$ that corresponds to $T=T_c$ is given, as a function of
$\beta$, by
\cite{vor3}
\be
L_t^c(\beta)  \simeq
\frac{1}{1.55} (\beta - 0.37).
\label{eqn_Ltcrit}
\ee
We see that, at both values of $\beta$, the lower value of
$T$ corresponds to $T \simeq 0.7 T_c $ while the
higher value corresponds to $T \simeq 1.4 T_c $.

For each of these lattices we calculate the average
action with a line of twisted plaquettes of length $r$ 
which introduces two $Z_2$ monopoles a distance $r$ apart as
discussed earlier. For each value of $r$ we perform
a Monte Carlo calculation, typically with $10^6$ sweeps, of 
the average total action, $\langle S(r)\rangle$. The
action here is
$S = \sum_p (1 - \frac{c_p}{2}TrU_p)$ where $U_p$ is the
product of the four SU(2) matrices around the plaquette $p$,
and $c_p$ is $-1$ for the twisted plaquettes and $+1$ otherwise. 
We plot in Tables \ref{table_l24b9} and 
\ref{table_l36b13.5} our results for the action. 
We turn now to a discussion of these results.

\subsection{${\bf T > T_c}$}
\label{subsec_highTcalc}

\begin{figure}[p]
\begin{center}
\epsfig{figure=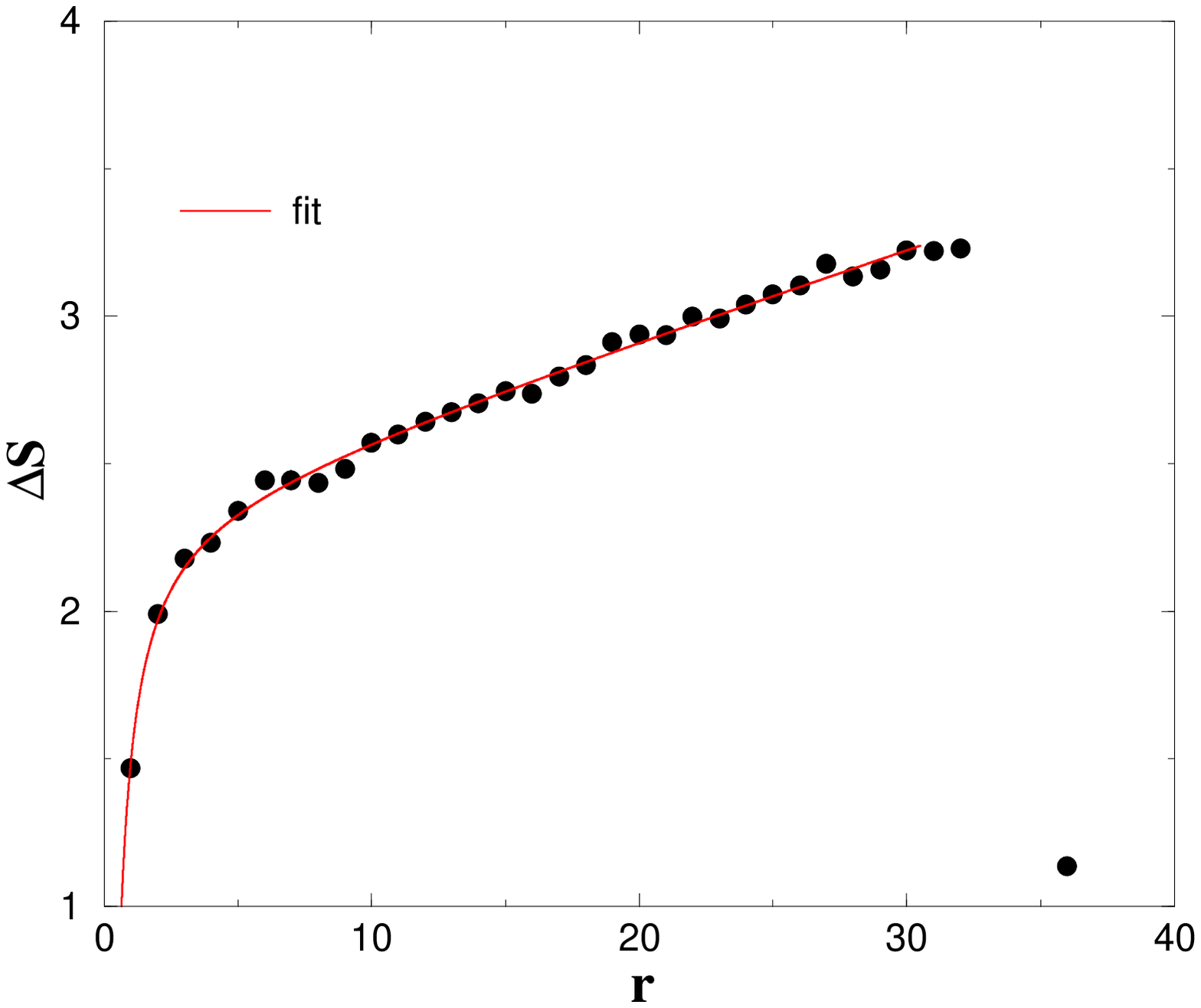, width=15cm} 
\end{center}
\caption[]{\label{fig_DS_l36t6}
{ The action of two $Z_2$ monopoles a spacelike distance $r$ apart 
on a $36^2 6$ lattice at $\beta=13.5$ for which $T > T_c$.}}
\end{figure}

In Fig.\ref{fig_DS_l36t6} we plot 
\be
\Delta S = \langle S(r) \rangle - \langle S(0) \rangle
\label{eqn_deltaS}
\ee
for the $\beta=13.5$ high-$T$ calculation. We observe 
an unambiguous confirmation of the expected linear
rise with $r$ at large $r$. Note that the point 
at $r=36$ has not been misplaced. Its value is given
by a periodic flux tube with no monopoles attached;
thus it is missing the large (divergent as $a \to 0$)
Coulombic self-energy of the monopoles, that otherwise 
contributes to $S(r)$ 
(except when $r$ is close to the lattice period).

We parameterise $\Delta S$ using a sum of linear and Yukawa terms
\be
\Delta S = c_0 + c {{e^{-\mu r}} \over {r}} +{\bar \sigma} r.
\label{eqn_yukawa_highT}
\ee
This conventional parameterisation is clearly approximate. We do not
expect the linear piece to be there at all at small values of $r$.
In addition, as remarked earlier, the Coulomb $1/r$ should
mutate into a $\log r$ dependence at larger $r$. Moreover
we expect a string/roughening correction to the linear term 
at large $r$, which can be calculated in perturbation theory
at high $T$
\cite{wall3}.
All of these corrections to eqn(\ref{eqn_yukawa_highT})
are small and we shall assume, perhaps optimistically, that 
they are negligible for our purposes.

We have performed fits with eqn(\ref{eqn_yukawa_highT})
to, for example, the range $r \in [1,30]$. We exclude $r=0$ from 
the fit for obvious reasons and we also exclude values $r > 30$ since
as $r$ approaches the lattice period, at some point it becomes
energetically favourable for the `Coulombic' part of the 
monopole-monopole flux to join through the periodic boundary
and for this to be matched by a periodic flux tube whose
location is now decoupled from the location of the twist.
We find that fits as in eqn(\ref{eqn_yukawa_highT}) possess a good
$\chi^2$ -- about 0.9 per degree of freedom for the best fit
shown in Fig.\ref{fig_DS_l36t6}.
There are two immediate questions: is the monopole string tension
as expected and what is the screening mass?

First, consider the monopole string tension. From the fit in
eqn(\ref{eqn_yukawa_highT}) we obtain a value 
\be
\bar \sigma = 0.0300 \pm 0.0009
\label{eqn_sigma_fit}
\ee
which is consistent with the value obtained from the $r=36$ 
value: $\bar \sigma = 0.0315(12)$. We can compare this with
our high-$T$ theoretical expectation in eqn(\ref{eqn_sigma}).
Plugging in $\beta=13.5$, $T/g^2=\beta/4L_t=13.5/24$, and
the value of $\alpha$ from eqn(\ref{eqn_alpha})
using the $L_t=6$ value of $\alpha_{pert}$ as listed
in Table 8 of 
\cite{wall3}
we obtain an expected value $\bar \sigma \simeq 0.035$.
If instead of employing the heuristic correction term
in eqn(\ref{eqn_alpha}) we use only the one-loop
value for $\alpha$ then we obtain  $\bar \sigma \simeq 0.024$.
Thus our fitted value of $\bar \sigma$ is in good agreement 
with our expectations; remarkably so given that the one-loop
perturbative expression has been 
derived for much higher $T$ than the values being
considered here.

A similar analysis of the $\beta=9$ calculation on a 
$24^2 4$ lattice yields best fits with a poorer,
albeit not unacceptable, $\chi^2$. We obtain a monopole
string tension $\bar \sigma = 0.078(3)$ which is
consistent with the string tension one finds with a periodic 
twist, $\bar \sigma = 0.0735(15)$. This can be compared 
with what one obtains from high-$T$ perturbation theory, 
eqns(\ref{eqn_sigma}): $\bar \sigma \simeq 0.056$ without 
the heuristic $O(g^2/T)$ correction in eqn(\ref{eqn_alpha}),
and  $\bar \sigma \simeq 0.081$ if we include it. Satisfactory
agreement, in other words.

We turn now to the screening mass $\mu$. Its fitted value is:
\be
\mu = 0.000 \pm 0.044 \ \ \ \ \ \ \ \ \ \beta=13.5
\label{eqn_mu_fit}
\ee
That is to say, we are quite consistent with a zero mass
and so a purely Coulombic (unscreened) short distance interaction 
between the monopoles. Certainly the screening mass is much less
than the $T=0$ mass gap. At high $T$ there are additional 
natural mass scales, the simplest being
$aT_c \simeq 0.12$, $aT = 1/6$, $\sqrt{g^2 T} \simeq 0.22$.
All of these are probably too large to be consistent with our 
value of $\mu$. There are other more dynamical mass scales such
as the (electric) Debye screening mass, $m_D$, and the
spatial string tension $\sigma_s$. We have performed
an explicit calculation of Polyakov loop correlations
on the same $36^2 6$ lattice at $\beta=13.5$ in order
to determine these masses
\cite{wall3}, 
and we find
\be
m_D = 0.103 \pm 0.005   \ \ \ \ \ \ \ \ \ \beta=13.5 
\label{eqn_mD_fit}
\ee
and
\be
\surd \sigma_s = 0.139 \pm 0.002 \ \ \ \ \ \ \ \ \ \beta=13.5.
\label{eqn_sigmas_fit}
\ee
The value of $\surd \sigma_s$ is too large for $\mu$ but
there is perhaps a possibility, at the $2\sigma$ level,
that the Debye screening mass provides this scale. 

A similar analysis of the $\beta=9$ calculation yields
\be
\mu = 0.2 \pm 0.2 \ \ \ \ \ \ \ \ \ \beta=9.0.
\label{eqn_mu_fit9}
\ee
This is again consistent with the (electric) Debye screening mass
that we have calculated on the same lattice at the same $\beta$,
\be
m_D = 0.16 \pm 0.01 \ \ \ \ \ \ \ \ \ \beta=9.0
\label{eqn_mD_fit9}
\ee
although, because of the larger errors, it is also consistent
with most of the high-$T$ scales referred to earlier (but
not to the $T=0$ mass gap).

The value of $\mu$ is hard to determine numerically
because, as we see from Fig.1, even at our relatively
large value of $\beta$ our resolution of the region in
which the `Coulombic' interaction dominates is poor.
We have tried fitting the values at small $r$ with a purely
Yukawa form, and that again produces a value consistent with 
zero. We have also modified the form in eqn(\ref{eqn_yukawa_highT}) 
by multiplying the linear piece by a factor $1-\exp(-\mu^\prime r)$.
This is quite natural; after all one would expect the
linear term to gradually disappear once $r$ becomes less than the
flux tube width. Such fits give values of $\mu$ and $\mu^\prime $  
that are non-zero and consistent with the value 
of $m_D$ in eqn(\ref{eqn_mD_fit}).

\subsection{${\bf T < T_c}$}
\label{subsec_lowTcalc}

\begin{figure}[p]
\begin{center}
\epsfig{figure=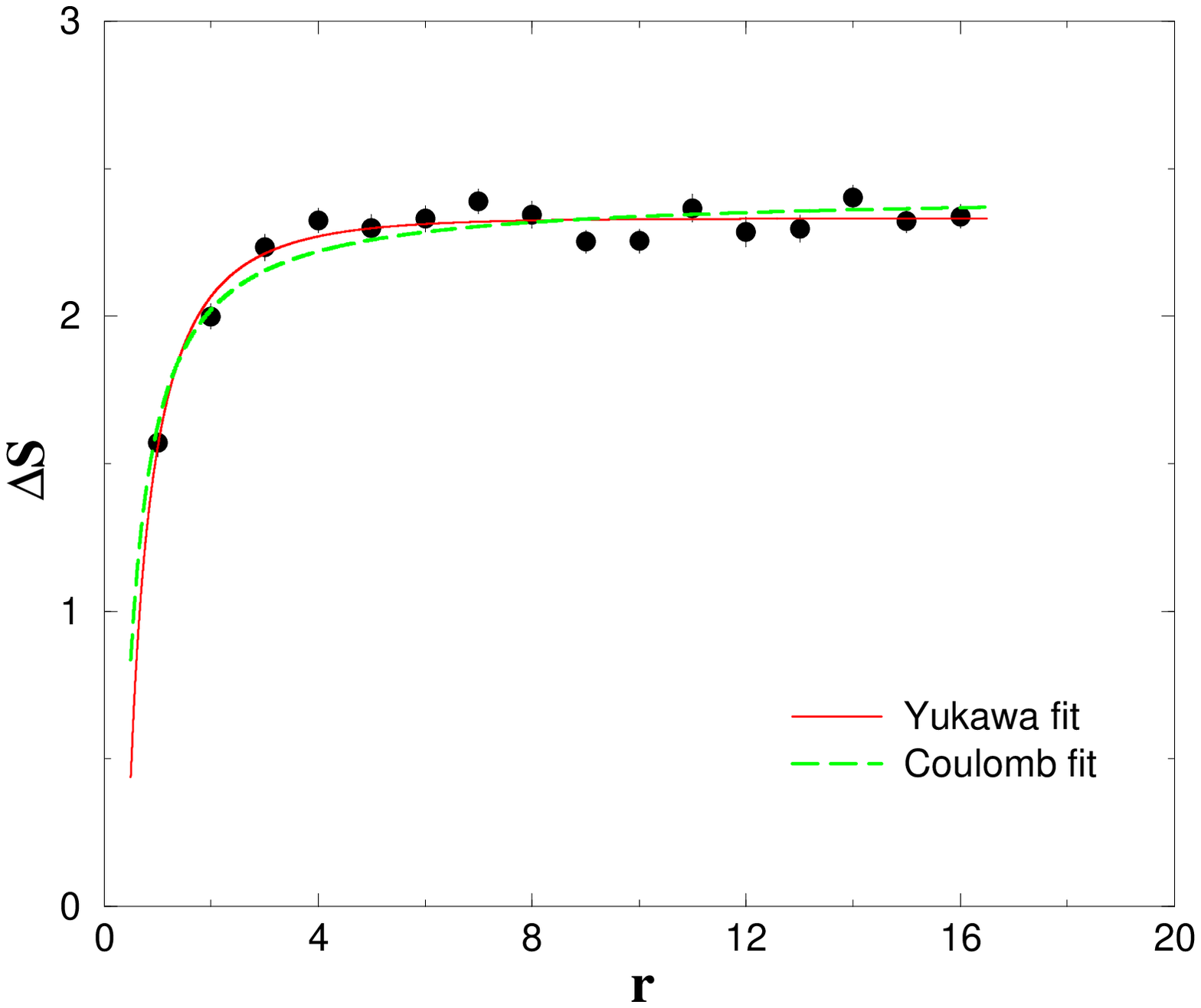, width=15cm} 
\end{center}
\caption[]{\label{fig_DS_l36t12}
{ The action of two $Z_2$ monopoles a spacelike distance $r$ apart 
on a $36^2 12$ lattice at $\beta=13.5$ for which $T < T_c$.}}
\end{figure}

In Fig.\ref{fig_DS_l36t12} we plot the action, $\Delta S$, 
of the two $Z_2$ monopoles as a function of their
separation, $r$, in the low-$T$ confining phase at
$\beta=13.5$. In contrast to the situation in the
deconfined high-$T$ phase, we see that there is no
linearly confining interaction between the monopoles.
This is as expected. spacelike $Z_2$ vortex tubes disorder 
timelike Wilson loops, and, in the confining phase,
will condense in the vacuum, thus costing no extra action.

At small separations the monopoles feel a Coulomb interaction
which should be screened at larger $r$ in this phase. Thus
we fit  $\Delta S$ to the Yukawa form:
\be
\Delta S = c_0 + c {{e^{-\mu r}} \over {r}}.
\label{eqn_yukawa_lowT}
\ee
We find that an unscreened Coulomb interaction is statistically
unacceptable (see the fits shown in Fig.\ref{fig_DS_l36t12}); 
there is clearly a non-zero screening mass
\be
\mu = 0.4 \pm 0.1  \ \ \ \ \ \ \ \ \ \beta=13.5 .
\label{eqn_mu_fitlowT}
\ee
Similar low-$T$ calculations on the $24^2 8$ lattice at $\beta=9$
also reveal a zero string tension and a non-zero screening
mass:
\be
\mu = 0.43 \pm 0.09   \ \ \ \ \ \ \ \ \ \beta=9.0 .
\label{eqn_mu_fitlowT9}
\ee
Since the above masses are in lattice units one would expect
the $\beta=13.5$ mass to be about two-thirds of the  $\beta=9$
value; which is quite possible within the rather large errors.
We also observe that taken together the values in
eqns(\ref{eqn_mu_fitlowT},\ref{eqn_mu_fitlowT9})
are not really compatible with the $T=0$ mass gap
\cite{spectrum3}:
$m_G \simeq 0.495(10)$ at $\beta=13.5$ and 
$m_G \simeq 0.76(1)$ at $\beta=9$. This is reminiscent
of what has been found 
\cite{mag4}
for the screening of the Abelian monopole flux 
in the maximally Abelian gauge in D=3+1. The screening mass
there was interpreted as the constituent gluon mass:
half the average mass of the lightest $0^+$
and $2^+$ glueballs. We note that our screening masses, in
eqns(\ref{eqn_mu_fitlowT},\ref{eqn_mu_fitlowT9}),
are consistent
\cite{spectrum3}
with being interpreted in this way.

\section{Conclusions}
\label{sec_conc}

We have shown that in the low-$T$ confining phase the interaction 
of a pair of $Z_2$ monopoles a distance $r$ apart 
can be described by a simple Yukawa interaction 
$\propto \frac{1}{r}\exp(-\mu r)$ where the effective screening 
mass $\mu$ is consistent with what one might expect
for the mass of a `constituent gluon'
\cite{mag4}.
In the high-$T$ deconfined phase
the $Z_2$ flux between widely separated monopoles is
collimated in a flux tube and the interaction energy increases
linearly with distance, i.e. the monopoles are `linearly
confined'. Such a spacelike flux tube, when closed upon
itself through a spatial boundary, becomes a domain wall
separating two high-$T$ $Z_2$ vacua. Its string tension
can be calculated in perturbation theory at very high $T$,
since the dimensionless coupling, $g^2/T$, becomes small there.
The string tension we obtain is surprisingly close to the 
value expected from one-loop perturbation theory, given that
in our `high-$T$' calculations $g^2/T$ is in fact close to unity.
At small $r$, where the flux tube has not yet formed, we find 
that the interaction is consistent with a Coulomb interaction 
possibly modified by very weak screening. This small screening
mass is just about consistent with the Debye electric
screening mass that we have also calculated.
All these calculations are for values of the lattice spacing
$a$ that are sufficiently small, expressed either in units of $g^2$ 
or in units of $T$, that we can expect negligible corrections
when taking the continuum limit. 

All our calculations have been for monopoles in 2+1 Euclidean 
dimensions where monopoles are instantons rather than solitons.
However we expect things to be very much the same in
3+1 Euclidean dimensions. At high $T$ there are, again,
perturbatively calculable  `domain walls' between 
differing $Z_2$ vacua, and these walls are just the
regions swept out by a periodic $Z_2$ flux tube  
as it moves through space-time in a spacelike direction. 
Consider two $Z_2$ monopoles joined by a Dirac string made 
invisible through the use of appropriately twisted plaquettes.
If their spacelike separation is large and they propagate
in a spacelike direction, the $Z_2$ flux between them
will generate a portion of such a $Z_2$ domain wall, that is
bounded by the monopole world lines. If we relabel the direction
of propagation as the time direction, so that the short 
Euclidean $t$ direction becomes a spatial direction, then
we can re-express the above as telling us that, within
this three dimensional space in which one of the spatial
directions is short (and equal to $1/T$) but the temperature
is now zero, the potential energy of two  $Z_2$ monopole sources 
is linearly confining.
(The asymmetric spatial geometry will by itself enforce 
confinement, if only logarithmic, at large $r$.) 
Moreover the corresponding string tension is calculable
in perturbation theory. At low $T$ by contrast, we expect
any such flux tubes to condense in the vacuum because the
system is confining, so that the monopoles experience
a screened Coulomb interaction just as in 2+1 dimensions.  

The aspect of the present calculations that could be readily
improved is that of the screening masses. This would
require calculations at smaller $a$ so as to increase the 
resolution of the region in which the Coulomb/Yukawa
potential is visible. If one were to consider not just
the total action, but the action over various distances
around each or both of the monopoles, the statistical
accuracy of the calculations would be greatly increased
as well.

\vskip 0.1in
\noindent{\bf Note added} As we were preparing to submit 
this paper for publication, a revised version of
\cite{rebbi}
appeared in which a linear fit to the high-$T$ inter-monopole
interaction is used. This was shown to be necessary in the recent,
more accurate, D=3+1 calculations of 
\cite{deforcrand},
and, as we have shown in this paper, the same is true in D=2+1.

\section*{Acknowledgments}
We are grateful to Chris Korthals Altes and Alex Kovner
for  stimulating discussions on issues related to those
dealt with in this paper. This collaboration began when one 
of the authors (ZS) was visiting Oxford and he is 
grateful to the Royal Society for funding the visit, 
under their European Science Exchange Programme. His work 
has also been supported in part by the Hungarian 
National Research Fund OTKA T032501 and the Bolyai J\'anos
Research Grant. AH and BL acknowledge their postdoctoral 
funding from PPARC.

\vfill

\eject

\newpage
\begin{table}
\begin{center}
\begin{tabular}{|c|c|c|}\hline
\multicolumn{3}{|c|}{$S \ ; \ \beta = 9$} \\ \hline
r & $24^2 4$ & $24^2 8$ \\ \hline
0 &     789.974(28) &     1583.412(34)    \\
1 &     791.492(26) &     1584.839(34)    \\
2 &     792.039(26) &     1585.299(34)    \\
3 &     792.205(25) &     1585.423(34)    \\
4 &     792.351(25) &     1585.536(29)    \\
5 &     792.443(29) &     1585.471(39)    \\
6 &     792.513(28) &     1585.542(36)    \\
7 &     792.640(28) &     1585.560(34)    \\
8 &     792.718(28) &     1585.488(36)    \\
9 &     792.825(23) &     1585.521(36)    \\
10 &     792.879(25) &    1585.578(41)    \\
11 &     792.948(26) &    1585.628(32)    \\
12 &     793.059(27) &    1585.507(37)    \\
13 &     793.191(24) &        \\
14 &     793.248(26) &        \\
15 &     793.302(25) &        \\
16 &     793.411(25) &        \\
17 &     793.429(25) &        \\
18 &     793.560(23) &        \\
19 &     793.579(25) &        \\
20 &     793.657(24) &        \\
21 &     793.701(18) &        \\
22 &     793.639(24) &        \\
23 &     793.223(24) &        \\
24 &     791.737(25) &        \\
\hline
\end{tabular}
\caption{\label{table_l24b9}
The total lattice action, $S$, on $24^2 4$ and $24^2 8$
lattices at $\beta = 9$, as a function of the separation,
$r$, between the two $Z_2$ monopoles.}
\end{center}
\end{table}

\begin{table}
\begin{center}
\begin{tabular}{|c|c|c|}\hline
\multicolumn{3}{|c|}{$S \ ; \ \beta = 13.5$} \\ \hline
r & $36^2 6$ & $36^2 12$ \\ \hline
   0    &    1760.398(29)    &     3522.752(39)    \\
   1    &    1761.865(32)    &     3524.321(49)    \\
   2    &    1762.390(28)    &     3524.750(42)    \\
   3    &    1762.577(30)    &     3524.983(44)    \\
   4    &    1762.630(27)    &     3525.077(41)    \\
   5    &    1762.739(31)    &     3525.050(44)    \\
   6    &    1762.841(30)    &     3525.081(45)    \\
   7    &    1762.843(33)    &     3525.140(41)    \\
   8    &    1762.832(31)    &     3525.095(45)    \\
   9    &    1762.882(24)    &     3525.004(38)    \\
   10    &    1762.969(30)    &     3525.006(40)    \\
   11    &    1762.998(27)    &     3525.118(47)    \\
   12    &    1763.041(29)    &     3525.038(49)    \\
   13    &    1763.072(30)    &     3525.049(45)    \\
   14    &    1763.102(28)    &     3525.152(41)    \\
   15    &    1763.144(31)    &     3525.074(39)    \\
   16    &    1763.134(26)    &     3525.088(41)    \\
   17    &    1763.195(31)    &    \\
   18    &    1763.234(28)    &    \\
   19    &    1763.311(30)    &    \\
   20    &    1763.336(29)    &    \\
   21    &    1763.335(29)    &    \\
   22    &    1763.397(29)    &    \\
   23    &    1763.390(28)    &    \\
   24    &    1763.437(29)    &    \\
   25    &    1763.472(27)    &    \\
   26    &    1763.501(28)    &    \\
   27    &    1763.576(29)    &    \\
   28    &    1763.533(32)    &    \\
   29    &    1763.557(30)    &    \\
   30    &    1763.622(28)    &    \\
   31    &    1763.619(30)    &    \\
   32    &    1763.629(32)    &    \\
   36    &    1761.537(28)    &     3522.863(45)    \\
\hline
\end{tabular}
\caption{\label{table_l36b13.5}
The total lattice action, $S$, on $36^2 6$ and $36^2 12$
lattices at $\beta = 13.5$, as a function of the separation,
$r$, between the two $Z_2$ monopoles.}
\end{center}
\end{table}



\begin{thebibliography}{99}

\bibitem{rebbi}
Ch. Hoelbling, C. Rebbi and V. Rubakov,
hep-lat/0003010.

\bibitem{vor3}
A. Hart, B. Lucini, Z. Schram and M. Teper,
JHEP 0006 (2000) 040.

\bibitem{vor4}
L. Del Debbio, A. Di Giacomo and B. Lucini,
hep-lat/0006028.

\bibitem{thooft}
G. `t Hooft,
Nucl. Phys. B138 (1978) 1; B153 (1979) 141.

\bibitem{wall3}
C. Korthals Altes, A. Michels, M. Stephanov and M. Teper,
Phys. Rev. D55 (1997) 1047.

\bibitem{deforcrand}
Ph. de Forcrand, M. D'Elia and M. Pepe,
hep-lat/0007034.

\bibitem{susskind}
M. Srednicki and L. Susskind,
Nucl. Phys. B179 (1981) 239.

\bibitem{spectrum3}
M. Teper,
Phys. Rev. D59 (1999) O14512.

\bibitem{Tc3}
M. Teper,
Phys. Lett. B313 (1993) 417.

\bibitem{mag4}
A. Hart and M. Teper,
Phys. Rev. D58 (1998) O14504.

\end{thebibliography}
\end{document}